\documentclass[twoside]{dis04}
\usepackage{amssymb,citesort,epsfig}
\def\runauthor{Stefan Kretzer}
\def\shorttitle{Strangeness Asymmetry}

\begin{document}

\title{Neutrino-Production of Charm and 
the Strangeness Asymmetry of the Nucleon
}

\author{Stefan Kretzer}

\address{Physics Department and
RIKEN-BNL Research Center,\\ 
Brookhaven National Laboratory,
Upton, New York 11973, USA}

\maketitle

\abstracts{Interest in the strange nucleon sea 
has been renewed when it was realized that
the  strangeness asymmetry $s^-=s-\bar{s}$ plays
a prominent role in the interpretation of the 
NuTeV weak mixing angle anomaly. I review the NLO QCD
calculation of the neutrino-production of opposite-sign 
dimuons as the experimental signature of the strange quark
parton density. Results from a recent CTEQ fit are presented 
and discussed with respect to their stability under NLO corrections
and their impact on the NuTeV measurement.}

\section{Introduction: Sea Quarks}
If the sea quarks of the nucleon could be
considered as ``resolved gluons'', 
they 
would inherit the gluon's flavour blindness and 
CP conjugation symmetry; i.e.
\begin{equation}\label{eq:naive}
\left. {\bar u}(x) \right|_{k_\perp^2 > \mu^2} =
\left.{\bar d}(x) \right|_{k_\perp^2 > \mu^2}=
\left.{\bar s}(x) \right|_{k_\perp^2 > \mu^2}=
\left. s(x) \right|_{k_\perp^2 > \mu^2}\ \ , 
\end{equation}
where the restriction on $k_\perp$ phase space 
generically denotes some perturbative cut-off.
For heavy quarks, it seems that the phenomenology 
of heavy quark
production works reasonably well under the assumption that
the heavy quark masses act as physical cut-offs in the
perturbative regime ($m_Q > \mu$). 
This is certainly not true for  
light quarks, however, where there will necessarily be
contributions from $k_\perp^2 < \mu^2$
that do not respect Eq.~(\ref{eq:naive}). 
It has been firmly established already that
\begin{equation}\label{eq:asymuds}
{\bar u}(x) \neq {\bar d}(x) \neq {\bar s}(x) 
\end{equation}
and it remains to be settled by which amount the 
strange sea quark and anti-quark
distributions differ:
\begin{equation}\label{eq:asyms}
s^-(x) \equiv (s-\bar{s})(x) \neq 0\ \ \ .
\end{equation}
Here and throughout I am avoiding the 
notion sometimes found in the literature 
of (flavour or CP) symmetry ``violation''; there is no symmetry 
breaking implied by Eqs.~(\ref{eq:asymuds}), (\ref{eq:asyms})
[e.g.~CP conjugation turns $s(x)$ into ${\bar s}_{\bar p}(x)$
-- the anti-strange sea of the anti-proton] and it 
would rather be a puzzle if these were 
exact equalities than inequalities to some degree. 
A broad literature on model calculations 
(see e.g.~\cite{models}) of the the sea quark 
boundary conditions (at $\mu$)
covers fascinating approaches to
non-perturbative dynamics ranging from light-cone wave functions 
over meson cloud models to the chiral quark soliton model. 
Here I restrict myself to the observation that 
the inequality (\ref{eq:asyms}) seems unavoidable 
and will look at data on neutrino-production of charm 
\begin{equation}\label{eq:charmcc}
\nu_\mu\ s\ \rightarrow c\ \mu^- \ \ \ \& 
\ \ \ {\bar \nu}_\mu\  {\bar s}\ \rightarrow {\bar c}\ \mu^+ 
\end{equation} 
to quantify if the amount can be significant. The experimental 
signature of the process (\ref{eq:charmcc}) are opposite sign 
dimuons (the second muon stemming from the charm decay)
in an active target \cite{CN};
I will next give an
overview of the corresponding QCD calculations.

\section{Neutrino-Production of Charm at NLO}
Chromodynamic corrections to the inclusive charm production process
in Eq.~(\ref{eq:charmcc}) 
were first calculated 
more than 20 years ago \cite{gotts}, a re-calculation e.g.~in \cite{gkr1}
fixes typos and provides modern ${\overline{\rm MS}}$
conventions which are also identical to 
the $m_s \rightarrow 0$ limit of the corresponding 
NLO corrections \cite{ks} in the ACOT scheme \cite{acotcc}.
In order to meet the real world experimental requirements 
of applying acceptance corrections to data \cite{CN}
taken with non-ideal detectors, differential NLO distributions
were calculated in \cite{gkr2} and \cite{disco} that provide
the charm hadron (D meson) kinematics in terms of the fragmentation
$z$ variable and rapidity $\eta$. The $d \sigma / dx dy dz d\eta$ 
code DISCO \cite{disco} exists as an interface to the NuTeV MC 
event generator.

For detailed NLO results I have to refer the reader
to the original articles listed above. 
In this short write-up I have to
restrict myself to an itemized summary: 
\begin{itemize}
\item[(i)] The NLO calculations all agree 
(some early discrepancies have been clarified).
\item[(ii)] For the fixed target kinematics under investigation, the NLO 
corrections to the LO process are modest, no bigger than 
${\cal{O}}(\lesssim 20\%)$.
\end{itemize}

\section{CTEQ Fit}
\begin{figure}[t]
\epsfig{figure=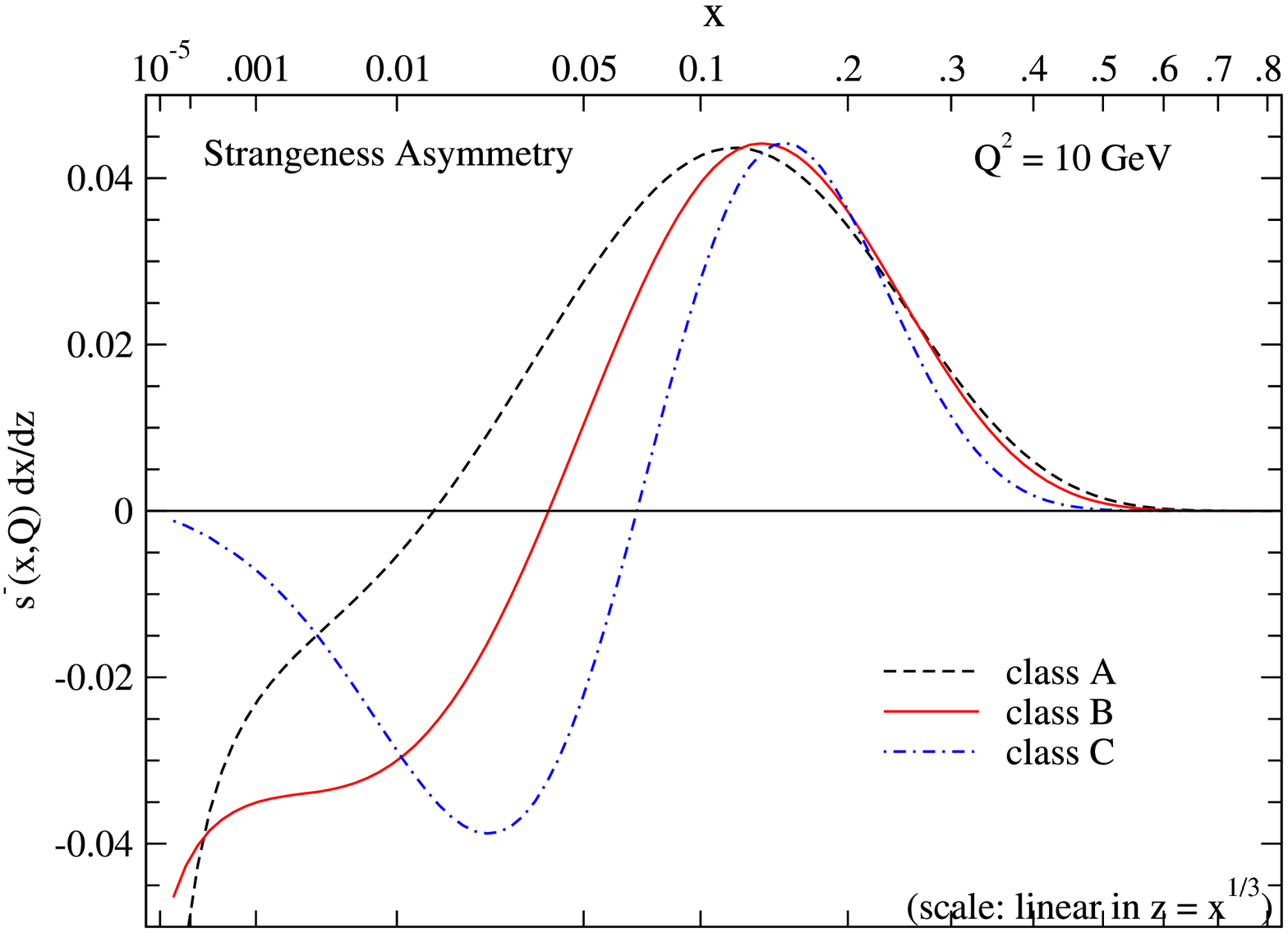,width=6cm}
\epsfig{figure=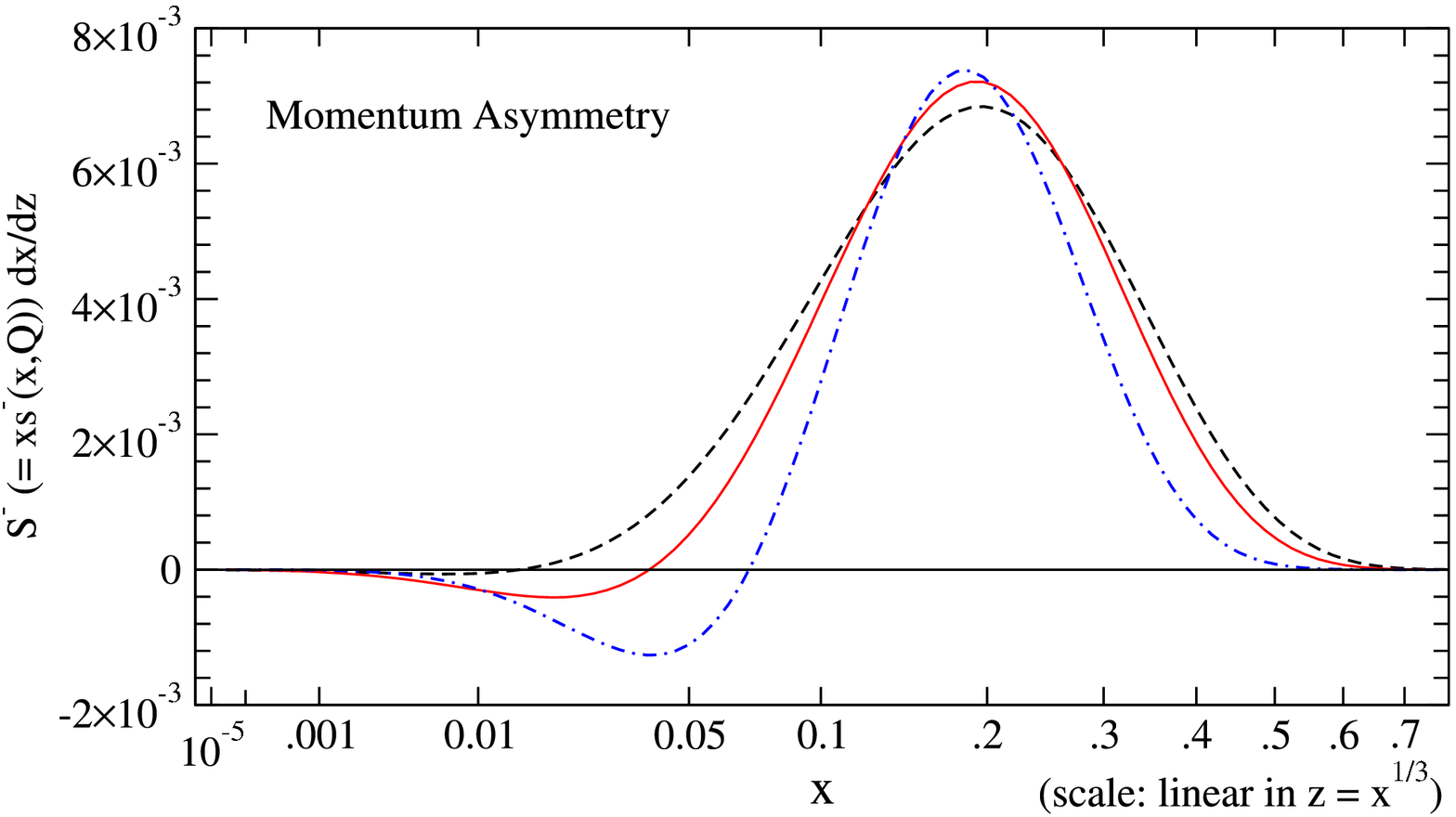,width=6cm}
\caption{Representative results of the CTEQ strangeness
asymmetry analysis.\label{fig:sasym}}
\end{figure}

Typical results of a recent CTEQ global data
analysis \cite{DimuonFitting} that includes
the dimuon data in \cite{CN} are shown in 
Fig.~\ref{fig:sasym}.
An essential constraint on these fits is the 
sum rule
\begin{equation}
\int \left[s(x)-\overline{s}(x)\right] dx = 0 \, ,
\label{eq:strnumsumrule}
\end{equation}
and a stable tendency of the fit is to realize the constraint
through a change of sign from negative to positive with
increasing $x$, resulting in a positive 
second moment integral
\begin{equation}
\left[ S^{-}\right] \equiv
\int x \left[ s(x)-{\bar{s}}(x)\right] dx \; .
\label{eq:mom2}
\end{equation}
Eq.~(\ref{eq:mom2}) is not overly sensitive to the low-$x$ ambiguities
visible in Fig.~\ref{fig:sasym} -- compare the number
asymmetry on the left plot with the momentum asymmetry on the right.
It is the second moment
(which is not among the local quark operators 
probed in DIS) that the NuTeV anomaly is mostly 
sensitive to, through an approximately linear 
relation between $\sin^{2}\theta_{W}$ and $[S^-]$
that was first derived in \cite{davidson}.

Note that the results in Fig.~\ref{fig:sasym}
have been obtained by a fit that neglects the 
NLO corrections discussed in the previous section
for consistency with the acceptance corrections that were applied
to the data \cite{CN} based on a LO model.
At worst, the CTEQ fit procedure constitutes a LO fit
with spurious higher order terms from the 
evolution and correlation with the global data that are otherwise
described to NLO accuracy. However,
we do find the results to be very stable under NLO corrections 
and the uncertainty limit on $[S^-]$ below is considerably broader 
than the NLO effects. Final NLO results
will have to await a certified update of the data 
\cite{CN} where acceptance effects are
corrected based on the NLO theory \cite{disco}.

At this conference, P.~Spentzouris for the NuTeV collaboration
has presented \cite{spentz} results from a fit that is based on the calculations
\cite{gkr1,disco} and uses the data \cite{CN} that are also included
in the CTEQ analysis. While the
results are within our limits (\ref{eq:bounds}) below, 
it remains to be understood why 
they display a qualitative preference for a
change of sign from positive to negative and, accordingly, a negative
$[S^-]$. The issue is currently investigated jointly by NuTeV and CTEQ.

\section{Impact on the NuTeV Anomaly}

By the Lagrangian multiplier method one finds
a central value $[S^-] \simeq 0.002$ and 
conservative bounds
\begin{equation}\label{eq:bounds}
-0.001 < [S^-] < 0.004\ \ \ .
\end{equation}
As described e.g.~in Ref.~\cite{davidson,nuprl} this translates into
a shift
\begin{equation}
-0.005 <  \delta (\sin^{2}\theta_{W}) < +0.001
\end{equation}
in $\sin^{2}\theta_{W}$ as measured in neutrino
scattering where there has been a 
$3\,\sigma$ discrepancy between the NuTeV result \cite{nuanom}
and the world average of other measurements of $\sin ^{2}\theta_{\mathrm{W}}$.
The shift in $\sin^{2}\theta_{W}$
corresponding to the central fit bridges a substantial part 
($\sim 1.5 \sigma$) of the original 
$3\,\sigma$ discrepancy.
For PDF sets with a shift toward the negative end, such as $-0.004$,
the discrepancy is reduced to less than 
$1\,\sigma $.  On the other hand, for PDF sets with a shift toward the
positive end, such as $+0.001$, the discrepancy remains. For
related discussions, see also the 
contributions \cite{spentz,others} to these proceedings.

\section{Conclusions}
Neutrino-production of charm is well understood in QCD and
it provides a direct handle on the strange sea asymmetry. 
This last undetermined asymmetry in the unpolarized quark sea
is bound to be non-zero but it is hard 
to quantify or even gauge for its significance in practice.
A model independent global parton structure analysis can
discriminate between models of non-perturbative strong interaction.
Recently, the observation was made 
that the non-perturbative effects may have to be  
disentangled from perturbative physics at the 3-loop level \cite{3loop}.
Apart from these interesting
issues in QCD phenomenology, 
limits on the second moment $[S^-]$ provide an essential 
systematic uncertainty in the NuTeV measurement of the weak mixing 
angle, which shows a $3 \sigma$ discrepancy with the 
standard model. The results of this study within their
uncertainty limits suggest that
the new dimuon data, the Weinberg angle measurement,
and other global data sets used in QCD parton structure
analysis can all be consistent within the standard model of
particle physics.

\section*{Acknowledgments}
The results presented here have been obtained in collaborations 
\cite{gkr1,gkr2,disco,nuprl,DimuonFitting,ks} 
with J.~Huston, M.~Gl\"{u}ck, H.L.~Lai, D.~Mason, P.~Nadolsky, 
F.~Olness, J.~Owens, J.~Pumplin, M.H.~Reno, E.~Reya, 
I.~Schienbein, D.~Stump and W.-K.~Tung.
Participation at DIS04 financed by 
RIKEN-BNL; work supported by DOE 
contract No.~DE-AC02-98CH10886 and RIKEN-BNL.


\begin{thebibliography}{99}
%
\bibitem{models}
S.J.~Brodsky, B.-Q.~Ma,
Phys.~Lett.~{\bf B381}, 317 (1996);
A.I.~Signal and A.W.~Thomas, Phys.~Lett.{\bf B191}, 205 (1987);
F.G.~Cao and A.I.~Signal,
Phys.~Lett.~{\bf B559}, 229 (2003);
M.~Wakamatsu, 
Phys.~Rev.~{\bf D} 67, 034005 (2003), 
and references therein.
%
\bibitem{CN}
CCFR and NuTeV Collab.~(M.~Goncharov {\it et al.}), Phys.~Rev.~{\bf D64},
112006 (2001); NuTeV Collab.~(M.~Tzanov {\it et al.}), {\tt
hep-ex/0306035}.
%
\bibitem{gotts}
T.~Gottschalk, Phys.\ Rev.\ D {\bf {23}} (1981) 56
%
\bibitem{gkr1}
M.~Gl\"{u}ck, S.~Kretzer and E.~Reya,
Phys.~Lett.~B {\bf 380}, 171 (1996); B {\bf 405}, 391 (1996) (E).
%
\bibitem{ks}
S.~Kretzer, I.~Schienbein,
Phys.~Rev.~{\bf D} 58, 094035 (1998).
%
\bibitem{acotcc}
M.A.G.~Aivazis, F.I.~Olness, W.-K.~Tung,
Phys.~Rev.~Lett.~65, 2339 (1990). 
%
\bibitem{gkr2}
M.~Gl\"{u}ck, S.~Kretzer and E.~Reya,
{\bf B} 398, 381 (1997); {\bf B} 405, 392 (1997) (E).
%
\bibitem{disco}
S.~Kretzer, D.~Mason and F.I.~Olness, 
Phys.~Rev.~{\bf D} 65, 074010 (2002).
%
\bibitem{DimuonFitting}
F.~Olness, J.~Pumplin, D.~Stump, J.~Huston, P.~Nadolsky, H.L.~Lai, S.~Kretzer, 
J.F.~Owens, W.K.~Tung, {\tt hep-ph/0312323}. 
%
\bibitem{davidson}
S.~Davidson, S.~Forte, P.~Gambino, N.~Rius, A.~Strumia,
JHEP {\bf 0202}, 037, 2002.
%
\bibitem{spentz}
P.~Spentzouris, these proceedings.
%
\bibitem{nuprl}
S.~Kretzer, F.~Olness, J.~Pumplin, M.H.~Reno and D.~Stump,
Phys.~Rev.~Lett.~93, 041802 (2004).
%
\bibitem{nuanom}
NuTeV Collaboration, G.P.~Zeller {\it et al.},
Phys.~Rev.Lett.~{\bf 88}, 091802  (2002).
%
\bibitem{others}
Contributions to these proceedings by 
K.~Diener,  S.~Kumano, and R.~Thorne.
%
\bibitem{3loop}
S.~Catani, D.~de Florian, G.~Rodrigo and W.~Vogelsang,
{\tt hep-ph/0404240}.
%
\end{thebibliography}
\end{document}